\documentclass[11pt]{article}
\usepackage[utf8]{inputenc}
\usepackage{natbib}
\usepackage[colorlinks=true, citecolor=blue,urlcolor=blue, linkcolor=blue]{hyperref}
\usepackage{graphicx}
\usepackage[tmargin=1in,bmargin=1in,lmargin=1in,rmargin=1in]{geometry}
\usepackage{amsmath}
\usepackage{xcolor}
\usepackage{subcaption}
\usepackage{float}
\usepackage{authblk}

\title{Examining Directional Strain Sensitivity of Shaped Optical Fibre Embedded in Polyurethane Strip}

\author[1]{Musab Al Hasani}
\author[1]{Guy Drijkoningen}
\author[2]{Thomas Reinsch}
\affil[1]{Department of Geoscience and Engineering, Delft University of Technology, Stevinweg 1, 2628 CN Delft, The
Netherlands}
\affil[2]{GFZ German Research Centre for Geosciences, Telegrafenberg
14473 Potsdam, Germany}

\begin{document}

\maketitle

\begin{abstract}
Distributed strain sensing using a straight optical-fibre cable suffers from a decreased strain sensitivity away from the fibre’s axis. In this study, the directional sensitivity is enhanced via sinusoidally shaping the fibre that is embedded in a polyurethane strip. First, the directional sensitivity is quantified via a relatively simple analytical model. Second, static-strain measurements using a Brillouin Optical Frequency Domain Analysis (BOFDA) system are carried out on a physical sample of the strip.  Two different directions are examined, namely in-line and broadside orientations.  The former involves deforming parallel to the plane where the fibre is sinusoidally embedded, whereas the latter means deforming the plane normal to it. It can be observed that the response to the deformation in these orientations is opposite, i.e. the fibre experiences negative strain, so shortening, for an in-line deformation, and a positive strain, so stretching, for a broadside deformation. It is found that the fibre is slightly more sensitive in the in-line direction which agrees with the behaviour as predicted by the model. Also, we see that the measured strain along the optical fibre is mainly influenced by the elastic properties of the embedding material due to, most importantly, the Poisson’s ratio as well as the geometrical parameters of the sinusoidally shaped fibre.
\end{abstract}
\section{Introduction}
Distributed fibre optical sensing is an emerging technology in the field of static and dynamic strain sensing. This technology has been utilised in wide range of applications including for monitoring the integrity of structures as well as for monitoring the strain and temperature for various settings. Distributed Strain Sensing (DSS) is an umbrella term used to describe a number of technologies to measure static and dynamic strains. Distributed Acoustic Sensing (DAS) or Distributed Vibration Sensing (DVS) are commonly used terms that refer to technologies that are developed to retrieve dynamic-strain measurements. In seismology, DAS/DVS has been adapted for passive and active seismic measurements. Early adoptions  were implemented for borehole monitoring through several Vertical Seismic Profiling (VSP) trails \citep{Mestayer2011, Daley2013, Frignet, Mateeva2014}. Even though the signal quality retrieved from DAS measurements still, for the most part, lags behind the quality of geophone measurements, DAS provides the opportunity to have denser spatial sampling with lower costs.   

Several implementations of DAS for surface seismic have been executed that include passive and active measurements. In the larger scale, DAS has been used for earthquake observations by \cite{Lindsey2017} and for imaging geological structures by \cite{Jousset2018}.  For near-surface velocity estimation,  Multichannel Analysis of Surface Waves (MASW) with passive and actives sources is carried out with DAS, as presented in \cite{Cole2018} as well as for site characterisation by \cite{Spica2020}. DAS also has been used to retrieve the reflection response at the surface using multiple shallow upholes \citep{Bakulin2017}. Using DAS via sensing with a straight fibre showed a directional sensitivity limitation as discussed in \cite{Mateeva2014}. This limitation manifests as a decreased sensitivity to broadside waves, i.e. waves perpendicular to the fibre. Helically wound fibres were introduced by  \cite{Kuvshinov2016} to enhance the broadside sensitivity. A patent by  \cite{Den} followed by introducing similar concepts. Attempts of implementing the helically wound fibres were presented in \cite{Hornman2017} and \cite{Urosevic2018}. The idea of shaping a fibre into a helix has been extended to so-called nested helices in theoretical studies, with the aim to estimate the full strain tensor, in \cite{LimChenNing2018} and \cite{Innanen2018}. 

\indent Although the fibre can be shaped to one's wishes, an implicit assumption in the latter two theoretical approaches is that the fibre needs to be embedded in a material in order to be manufactured as a cable. So the embedding material needs to be part of the model and the embedding material will affect the behaviour of the fibre. Here we address this issue. 

\indent In our study, we adapted the sinusoidally shaped fibre concept by embedding it within a strip of polyurethane, in our case made of a type called Conathane\textsuperscript{\textregistered}. We first provide an analytical description of the strain experienced by the embedded fibre.  This is followed by a discussion of experiments conducted with a Brillouin Optical Frequency Domain Analysis (BOFDA) system examining the difference in sensitivity between the two sides of the strip to strain. And finally we aim to model the results and match it with the observed data, to see whether the behaviour of the fibre embedded in the strip is well described by the model.

\section{Deforming a Sinusoidally Shaped Fibre with a Static Load }
\begin{figure}
\centering
\begin{subfigure}[b]{0.55\textwidth}
  \includegraphics[width=1\linewidth]{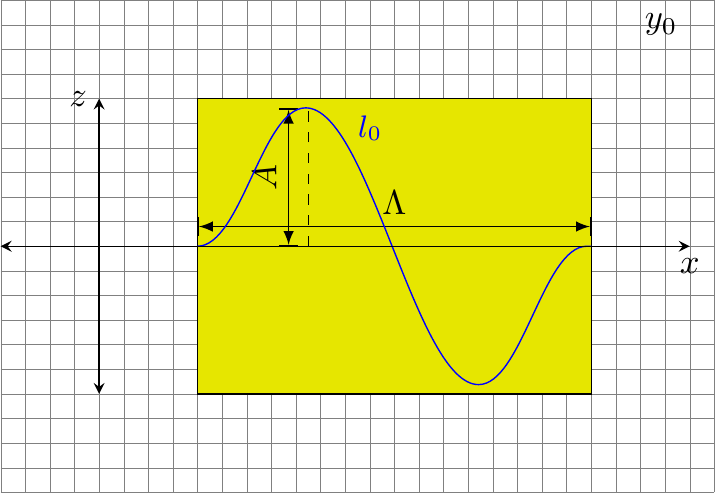}
  \caption{}
  \label{fig:sinus_no_def} 
\end{subfigure}

\begin{subfigure}[b]{0.55\textwidth}
  \includegraphics[width=1\linewidth]{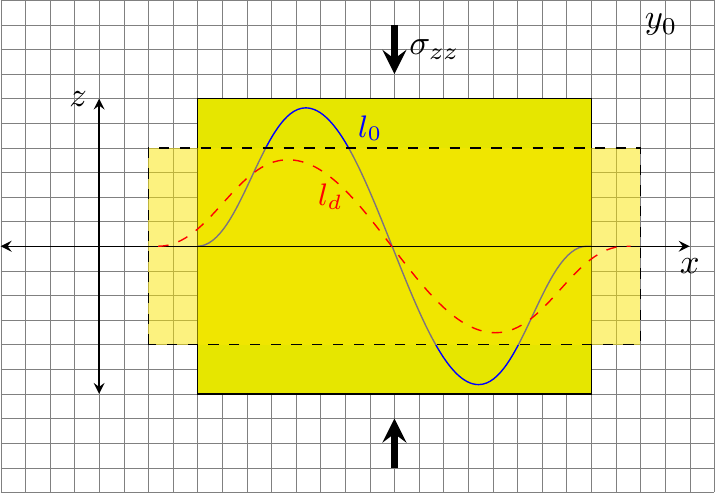}
  \caption{}
  \label{fig:sinus_IL} 
\end{subfigure}

\begin{subfigure}[b]{0.55\textwidth}
  \includegraphics[width=1\linewidth]{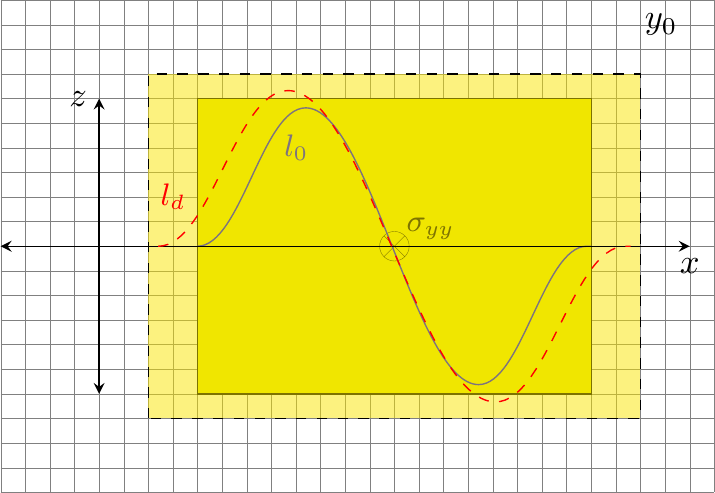}
  \caption{}
  \label{fig:sinus_BS} 
\end{subfigure}

\caption{\textit{Deformation of sinusoid embedded in a strip (a) before deformation, and after (b) in-line deformation and (c) broadside deformation.}}\label{fig:sinus}
\end{figure}

 A simple analytical model is derived for the strain experienced by a sinusoidally shaped fibre. Consider a sinusoid with a length denoted by $l_0$, embedded in a strip. A cross-section through the plane $y_0$ in which the sinusoidal fibre is located, is shown in figure \ref{fig:sinus_no_def}. In this model the parameters $A$ and $\Lambda$ are introduced that denote the amplitude and the wavelength of the sinusoid, respectively. 

 Two scenarios of deformation are considered here. The first one, that we call in-line deformation $\varepsilon_{zz}$, is due to a plane stress $\sigma_{zz}$ acting in the $z$-direction, as indicated in figure \ref{fig:sinus_IL}. The other is a broadside deformation $\varepsilon_{yy}$, where a plane stress $\sigma_{yy}$ is applied in the $y$-direction, as indicated in figure \ref{fig:sinus_BS}.

First we need to calculate the original undeformed fibre length $l_0$ since the measurement device measures along the fibre, not along the cable. If the shape is expressed by the function $f_0=A \sin(2\pi \frac{x}{\Lambda})$, then the length $l_0$ of one sinusoid is given by the arc length as 
\begin{equation}\label{eq:Lo_int}
	l_0 = \int_0^{\Lambda} \sqrt{1+4\pi^2 \frac{A^2}{\Lambda^2} \cos^2\left(2 \pi \frac {x}{\Lambda}\right)}\ dx.
\end{equation}

\subsection*{In-line Deformation}

\indent Deforming the strip with a compressive stress $\sigma_{zz}$ will result in a change in the shape of embedded sinusoid (see fig. \ref{fig:sinus_IL}). We assume that the stresses applied are low and within the linear elastic regime. Therefore, its relationship to the strain in the same direction, $\varepsilon_{zz}$, is  
\begin{equation}\label{eq:E}
    \sigma_{zz} = E_c \ \varepsilon_{zz},
\end{equation}
where $E_c$ is Young's modulus of the embedding cable. We assume that the displacement change of the fibre in $z$-direction is proportional to $\varepsilon_{zz}$ of the cable material, and in the $x$-direction by $\varepsilon_{xx} = - \nu_c \ \varepsilon_{zz}$, where $\nu_c$ is Poisson's ratio of the cable material. This deformation will cause $A$ to decrease due to the negative $\varepsilon_{zz}$ and $\Lambda$ to increase.
Following the assumption of plane strain and that the fibre is perfectly coupled in the $y_0$-plane, the out-of-plane strain  $\varepsilon_{yy}$ is assumed to be zero. 
This will result in change in the fibre shape as 

\begin{equation}\label{eq:f_d}
    f_d=\left(1+\varepsilon_{zz}\right) A \sin\left(2\pi \frac{x}{\Lambda(1- \nu_c\varepsilon_{zz})}\right).
\end{equation}

\noindent
Therefore, the length of the deformed fibre is found by the following equation: 

\begin{equation}\label{eq:Ld3_int}
	l_d = \int_0^{\Lambda(1 - \nu_c\varepsilon_{zz})}
	    \sqrt{
	       1 +  4 \pi^2 
	       \frac{ A^2 ( 1 + \varepsilon_{zz} )^2}
	            { \Lambda^2 (1- \nu_c\varepsilon_{zz})^2 }
	           \cos^2 \left( 
	                     2 \pi \frac{ x }{ \Lambda(1- \nu_c\varepsilon_{zz}) } 
	                  \right) 
	           } \ dx.
\end{equation}

\subsection*{Broadside Deformation}

In this case,  a plane stress of $\sigma_{yy}$ is applied to the strip causing the deformation of the sinusoid in the $x$- and $z$-directions as shown in figure \ref{fig:sinus_BS}. For a compressive stress, the sample will be shortened in the $y$-direction, so $\varepsilon_{yy} < 0$, but will be lengthened in the other directions, due to Poisson's ratio of the material. This will result in a deformed fibre shape $f_d$, expressed as 

\begin{equation}\label{eq:f_d2}
    f_d=\left(1 - \nu_c \varepsilon_{yy}\right) A \sin\left(2\pi \frac{x}{\Lambda(1 - \nu_c \varepsilon_{yy} )}\right).
\end{equation}
Thus, the length of the deformed fibre due to a broadside load is found as 

\begin{equation}\label{eq:Ld3_int2}
	l_d = \int_0^{\Lambda(1 - \nu_c\varepsilon_{yy})}
	    \sqrt{
	       1 +  4 \pi^2 
	       \frac{ A^2 }
	            { \Lambda^2  }
	           \cos^2 \left( 
	                     2 \pi \frac{ x }{ \Lambda(1 - \nu_c \varepsilon_{yy}) } 
	                  \right) 
	           } \ dx.
\end{equation}

\noindent
Since $\varepsilon_{yy}$ is negative for a compressive stress, an elongation in the $x$- and $z$-direction is expected. 

\subsection*{Fibre Strains and Sensitivities }

The strain of the traced fibre, i.e. $\varepsilon_{f}$,  assumes that the fibre is perfectly coupled to the embedding material. It can be described by the following expression: 

\begin{equation}\label{eq:ec}
    \varepsilon_{f} = \left ( \frac{l_d(\varepsilon_{xx},\varepsilon_{yy},\varepsilon_{zz})}
         {l_0} - 1 \right )
\end{equation}

\begin{figure} [H]
\captionsetup[subfigure]{labelformat=empty}
    \begin{subfigure}{.5\textwidth}
        \centering
        \includegraphics[width=\textwidth]{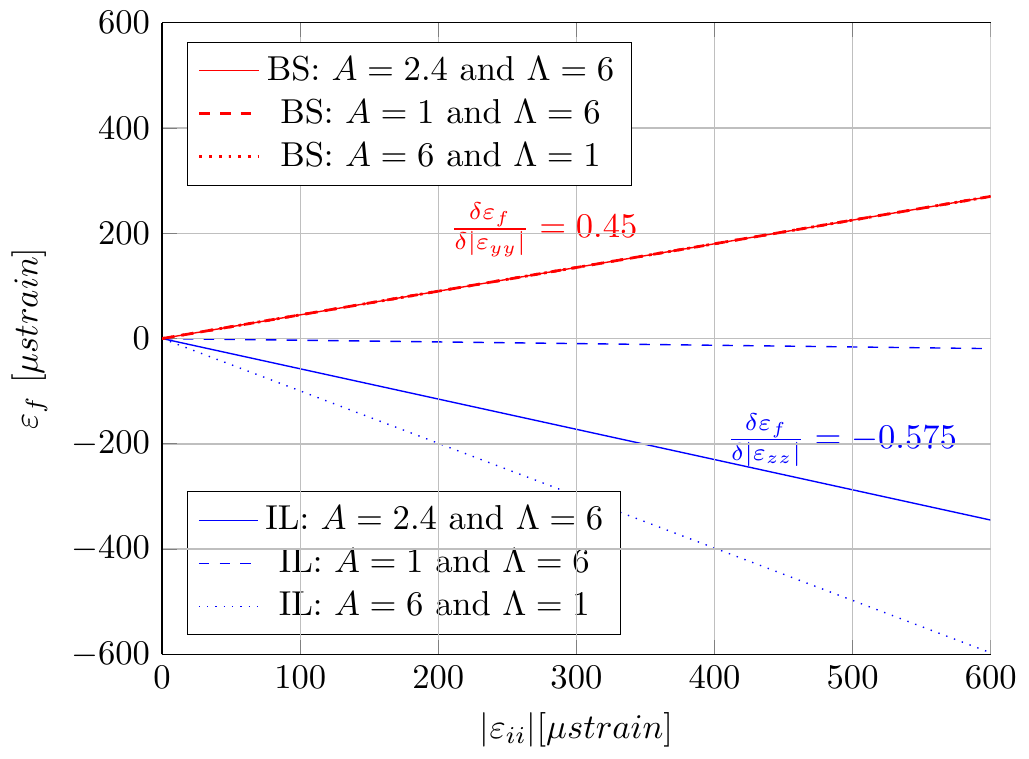}
        \caption{(a)}
        \label{fig:eps_f_vs_geom}
    \end{subfigure}%
    \begin{subfigure}{.5\textwidth}
        \centering
        \includegraphics[width=\textwidth]{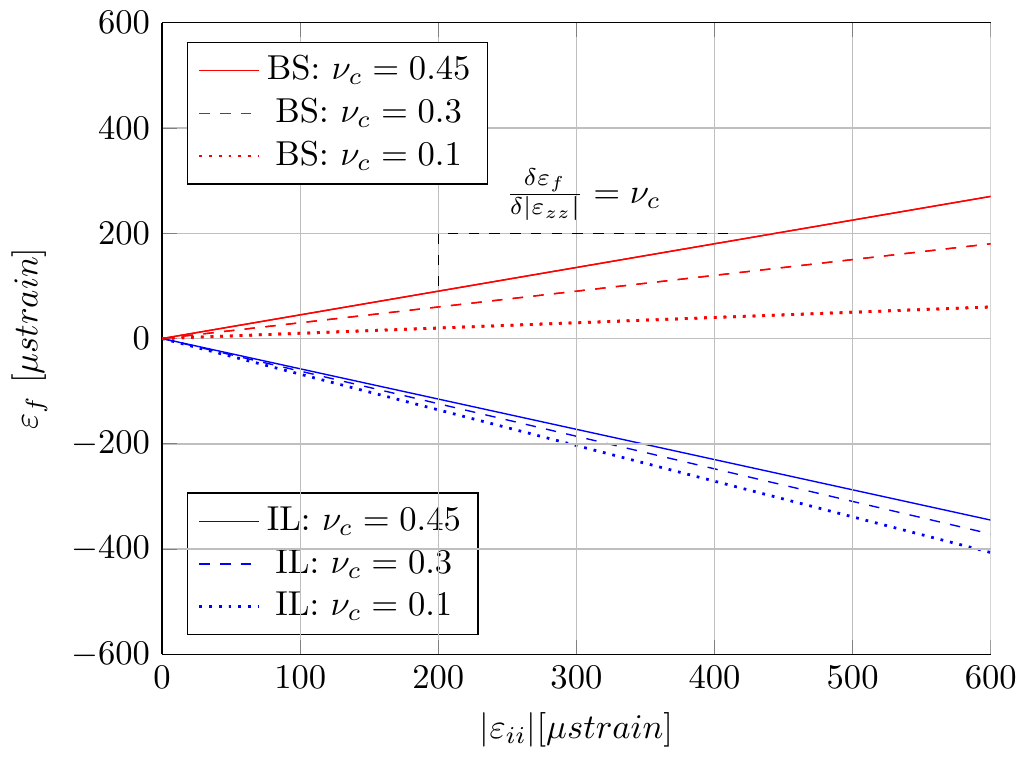}
        \caption{(b)}
        \label{fig:eps_f_vs_nu_c}
    \end{subfigure}%
    
\caption{\textit{Along-fibre strain $\varepsilon_f$ as function of in-line (IL) strain $\varepsilon_{zz}$ and broadside (BS) strain $\varepsilon_{yy}$ for (a) geometrical parameters $A$ and $\Lambda$ (in $cm$) with $\nu_c = 0.45$ and (b) various values of $\nu_c$, with $A = 2.4$ cm and $\Lambda = 6$ cm.}}
\label{fig:eps_f_vs_geom_vs_nu_c}
\end{figure}

To quantify the difference in sensitivity of the two scenarios, we evaluate $\varepsilon_f$ for both cases using expression \ref{eq:ec}. A comparison for the effect of the geometrical parameters of the fibre (i.e. $A$ and $\Lambda$) and the Poisson's ratio of the cable material $\nu_c$ is illustrated in figure \ref{fig:eps_f_vs_geom_vs_nu_c}.
It can be seen that the relations are as good as linear, due to $\varepsilon_{ii}$ being very small. The negative slope of the in-line scenario due to a load predicts a shortening of the fibre, unlike the elongation in case of the broadside load. This model predicts that the fibre is more sensitive in the in-line direction, as expected.

As for determining the sensitivity to the geometrical parameters, figure \ref{fig:eps_f_vs_geom} shows  the change for the traced fibre against strain due to the applied load, i. e. $\varepsilon_{zz}$ or $\varepsilon_{yy}$ for a given material with $v_c = 0.45$, which is within the range of Poisson's ratio of the examined Polyurethane strip, and various geometrical parameters $A$ and $\Lambda$.  From figure \ref{fig:eps_f_vs_geom}, it can be seen that for the in-line scenario, $\varepsilon_f$ is highly dependent on the geometrical parameters.  We see that for larger $A$ and smaller $\Lambda$, the in-line sensitivity is increased, and vice versa. 
This is unlike the broadside sensitivity  which is not affected by changing the geometry at all (fig. \ref{fig:eps_f_vs_geom}): the slope of the line always gives the Poisson ratio itself. This is a convenient characteristic. Then in figure \ref{fig:eps_f_vs_nu_c} the Poisson ratio is varied and it can be observed that the effect of Poisson's ratio is more pronounced on the broadside scenario as it is equal to the slope of the line. Although it is also affecting the in-line scenario, its effect is smaller.

\section{Static strain experiment with Brillouin Optical Frequency Domain Analysis (BOFDA)}

\noindent

\subsection*{Principle of Stimulated Brillouin Scattering}
Stimulated BOFDA makes use of the scattering caused by the interaction of incident light with vibrations in the molecules of the optical fibre. These vibrations are the result of an excited collection of molecules called acoustic phonons. This interaction is called Brillouin scattering and occurs in the 10-30 GHz range \citep{hartog2017}. This scattering process is not elastic (as used in particle physics)
because the energy of the incidence photon, and in turn its frequency, is altered. 

In our experiment we used a system based on this principle, made by the company fibrisTerre, the type fTB 2505.  It requires both ends of the fibre to be connected to the instrument, where a pump wave 
and a probe wave  (sometimes referred to as a Stokes wave) are injected simultaneously, by means of a sweep of frequencies. A fibre will resonate at a beat frequency in the response to these waves. That beat frequency will result in propagating fluctuations of density. The interaction between the waves and this fluctuation will cause a frequency shift in the pump wave, called Brillouin frequency-shift $f_B$. This is what is measured by the instrument. Further details of the method can be found in \cite{Nother2010}. 

Stretching or lengthening the fibre will increase $f_B$, whereas shortening the fibre will decrease $f_B$.  The strain experienced by the fibre can be evaluated using the following expression, as given in the fTB2502 manual: 
\begin{equation}
    \varepsilon_f = \frac{\Delta f_B }{C_{\varepsilon} } = \frac{f_{B,m}- f_{B, r}}{C_{\varepsilon}} 
    \label{eq:strain}
\end{equation}
where \\
\indent $\varepsilon_f$ \space \space \space fibre strain in $\mu \varepsilon$, \\
\indent $f_{B,m}$  measured Brillouin frequency shift in MHz, \\
\indent $f_{B,r}$ \space reference Brillouin frequency shift in MHz, \\
\indent $ C_{\varepsilon}$ \space \space \space strain coefficient of fibre in MHz/$\mu \varepsilon$. \\
\noindent $C_{\varepsilon}$  varies typically between $40$ to $50$ MHz/1000$\mu \varepsilon$ for different types of fibre, according to the fTB2502 manual. A value of $61$ MHz/1000$\mu \varepsilon$ is also reported by \cite{Nother2010}.

\subsection*{The Sample}

In order to verify whether the model describes/explains the behaviour of a real sample, company De Regt produced a sinusoidally shaped fibre embedded in a Conathane\textsuperscript{\textregistered} strip. The fibre is bend-insensitive with core/cladding diameter of 6.2/80$\mu m$ and a dual-acrylate coating of 170$\mu m$. The elastic properties of our optical fibre was not measured, typical values are contained in table \ref{tab:matProp} that are based on \cite{Pulker1982} and \cite{Sudheer2015}. The table shows the significant difference of elastic properties of fibre compared to the elastic properties of Conathane\textsuperscript{\textregistered}. The sample is 200 cm long, 6 cm wide and its thickness varies between 1.6 to 1.8 cm. This thickness variation could be attributed to the manual production process. The length of the fibre embedded within the Conathane\textsuperscript{\textregistered} strip is about 4 m long with extra fibre at both ends for the terminations.

Since the material properties of Conathane\textsuperscript{\textregistered} are not well known, a separate small experiment was performed on a cylindrical sample to determine Young's modulus $E_c$ and Poisson's ratio $\nu_c$. The strains to that cylindrical sample were of the order of 10 m$\varepsilon$'s. In contrast,  these are much lower than the strains applied to the sample with the sinusoidally shaped fibre which were in the order of 0.1 m$\varepsilon$. The values obtained from this small experiment are included in Table \ref{tab:matProp}. The Young's modulus and Poisson's ratio of Conathane\textsuperscript{\textregistered} are measured through a unaxial compressional test using MTS 815 system at a strain of 1\%. Note that this system is made for rock mechanics applications, hence it is not the best choice for rubber-like materials. This rises uncertainty as the applied strains $\varepsilon_{ii}$ to the fibre-containing sample are much lower than 1\% strain.

\indent 
\begin{table}[H]
\centering
\begin{tabular}{ |p{5cm}||p{3cm} |p{5cm}| }
 \hline
 \textbf{Material Property} & \textbf{Conathane\textsuperscript{\textregistered}}                                                            &      \textbf{Optical fibre}   \\\hline 
 Young's Modulus E (GPa)      &  0.0165 - 0.0175  &  16.56  \citep{Pulker1982}    \\ \hline
 Poisson's Ratio $\nu$ (l/l)  &  0.43 - 0.48      &  0.22 \citep{Sudheer2015}     \\ \hline
\end{tabular}
\caption{\textit{Elastic properties of Conathane\textsuperscript{\textregistered} and of optical fibre (at room temperature).}}
\label{tab:matProp}
\end{table}

\begin{figure}[!ht]
    \centering
    \includegraphics [width=0.9\textwidth] {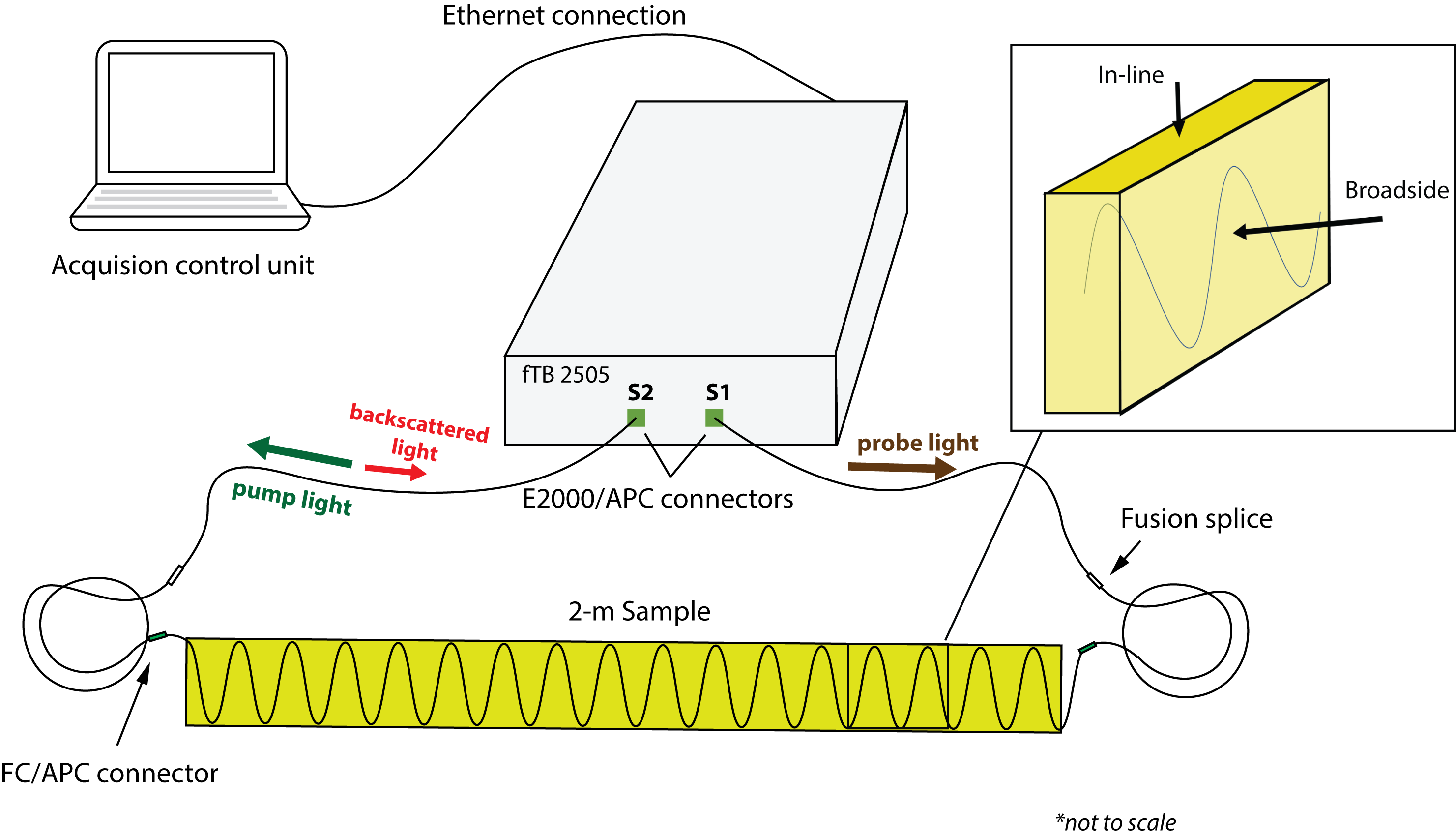}
    \caption{\textit{Set-up of static strain experiment with BOFDA. Note that the loads were applied in in-line and broadside orientations as illustrated. }}
    \label{fig:Exp_setup}
\end{figure}

\subsection*{Set-up of Experiment}
 
\begin{figure}[!ht]
    \centering
    \includegraphics [width=0.9\textwidth] {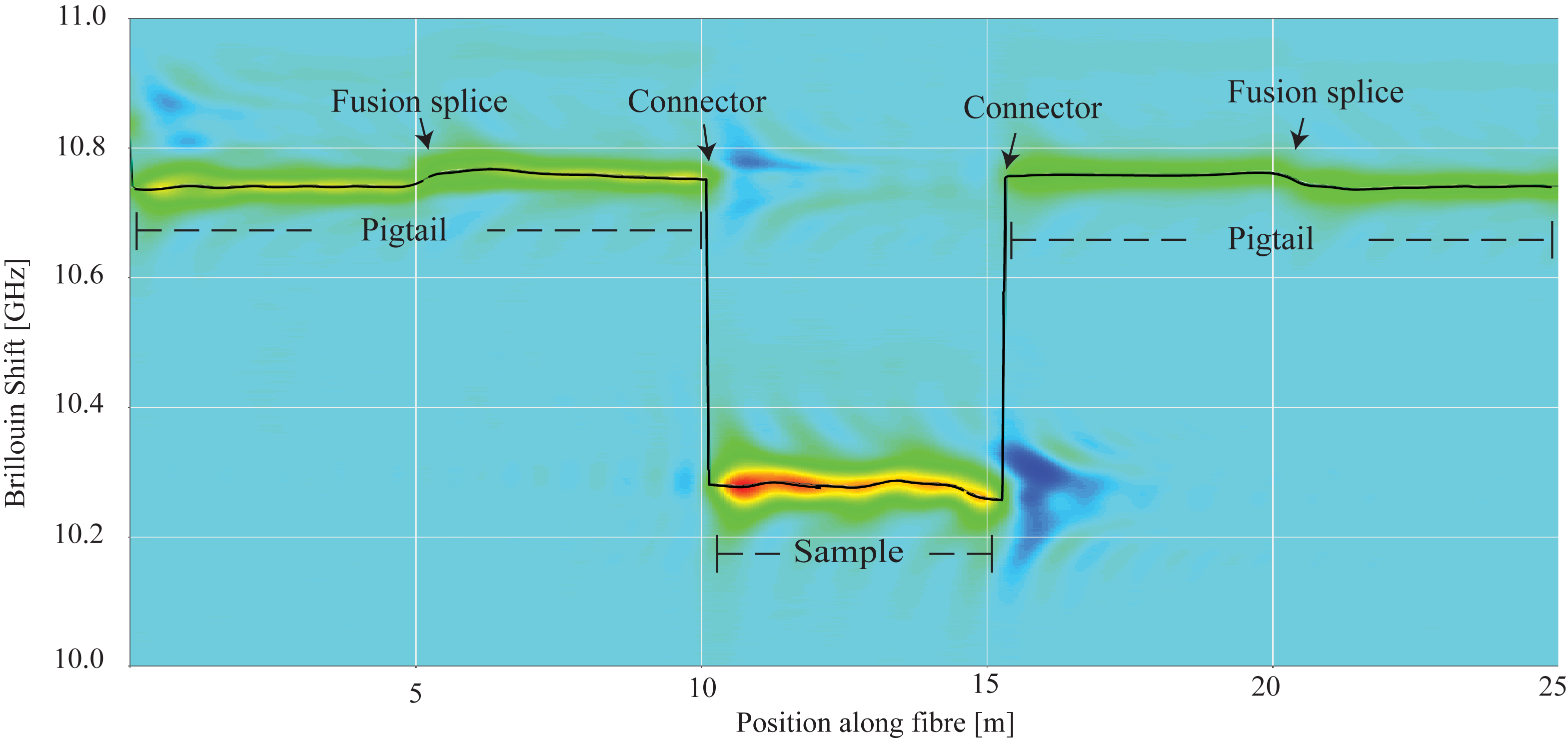}
    \caption{\textit{Reference spectrum of Brillouin frequency shift ($f_B$). Note that the sample includes extra fibre at both ends for the terminations. 
    }}
    \label{fig:BFS_spectrum}
\end{figure}

The set-up of the measurement is shown in figure \ref{fig:Exp_setup}. 
Two pigtails were connected to the sample with FC/APC connectors at the terminations of both ends. Both pigtails have fusion splices in the middle as marked in the figure. The ends of the pigtails are connected to the device with E2000/APC connectors. To optimise the acquisition process, a initial measurement was taken to choose the frequency range of interest.  The main parameters to optimise are the Brillouin Frequency Shift scan range and step as well as spatial resolution; a scan range of 10 to 11 GHz was chosen, with a step of 0.1 MHz (i.e. this is equivalent to steps of $2 \mu \varepsilon$ for $C_{\varepsilon}$  = 50 MHz/1000 $\mu \varepsilon$) for every 0.2 m, respectively. 
Figure \ref{fig:BFS_spectrum} shows the initial $f_B$ spectrum, annotated with the main sections of the fibre.  Based on this, the Brillouin-frequency scan range was limited to 10.2-10.4 GHz to reduce the frequency-sweeping time for the high-resolution measurement of 0.1 MHz to only include the fibre in the sample. 

\subsection*{Observed frequency shift due to In-line (\texorpdfstring{$z$}{z}) and Broadside (\texorpdfstring{$y$}{y}) loads}
\begin{figure}
    \centering
    \includegraphics[width=\textwidth]{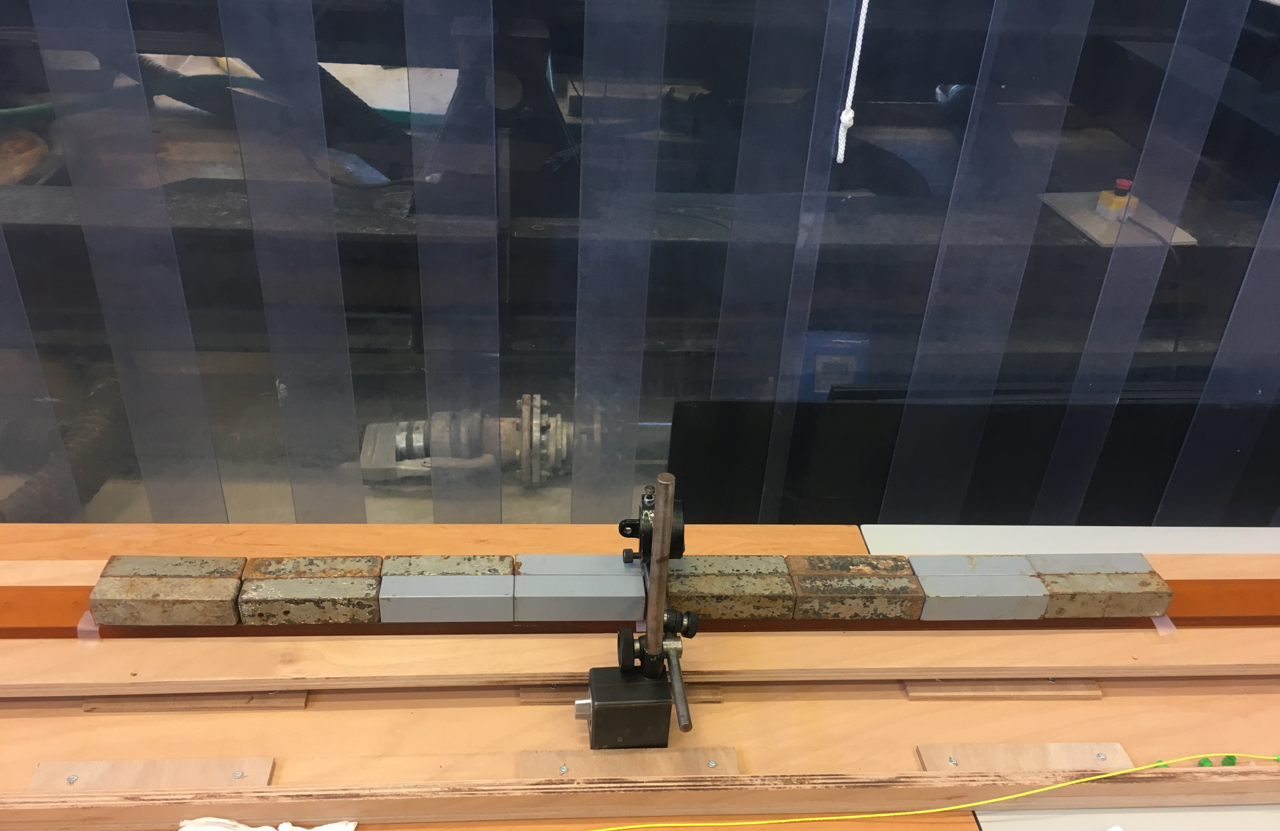}
    \caption{Broadside load over 1m section of the Conathane\textsuperscript{\textregistered} strip.}
    \label{fig:load}
\end{figure}

To investigate the directional sensitivity our sample, a range of loads was applied to the sample in in-line and broadside orientations. Here we define in-line to be the $z$-direction of the sinusoid and broadside to be the direction normal to the $xz$-plane, as in figure \ref{fig:sinus}. The loads were applied via weights that were uniformly put on a 1-m section in the middle of the strip. The loads were chosen such that they resulted in the same stress range, i.e. from 0 to 9000 Pa for both orientations. The stress $\sigma$ is derived from these loads using $\sigma = mg/A_c$,  where $m$ is  the  mass, $g$ the  gravity constant $g= 9.81\, \mbox{m/s}^2$, and $A_c$ the surface area of the cable to which the load was applied. An example of a load applied to a part of the strip is showing in figure \ref{fig:load}. 

\begin{figure} [H]
\captionsetup[subfigure]{labelformat=empty}
    \begin{subfigure}{.5\textwidth}
        \centering
        \includegraphics[width=0.9\textwidth]{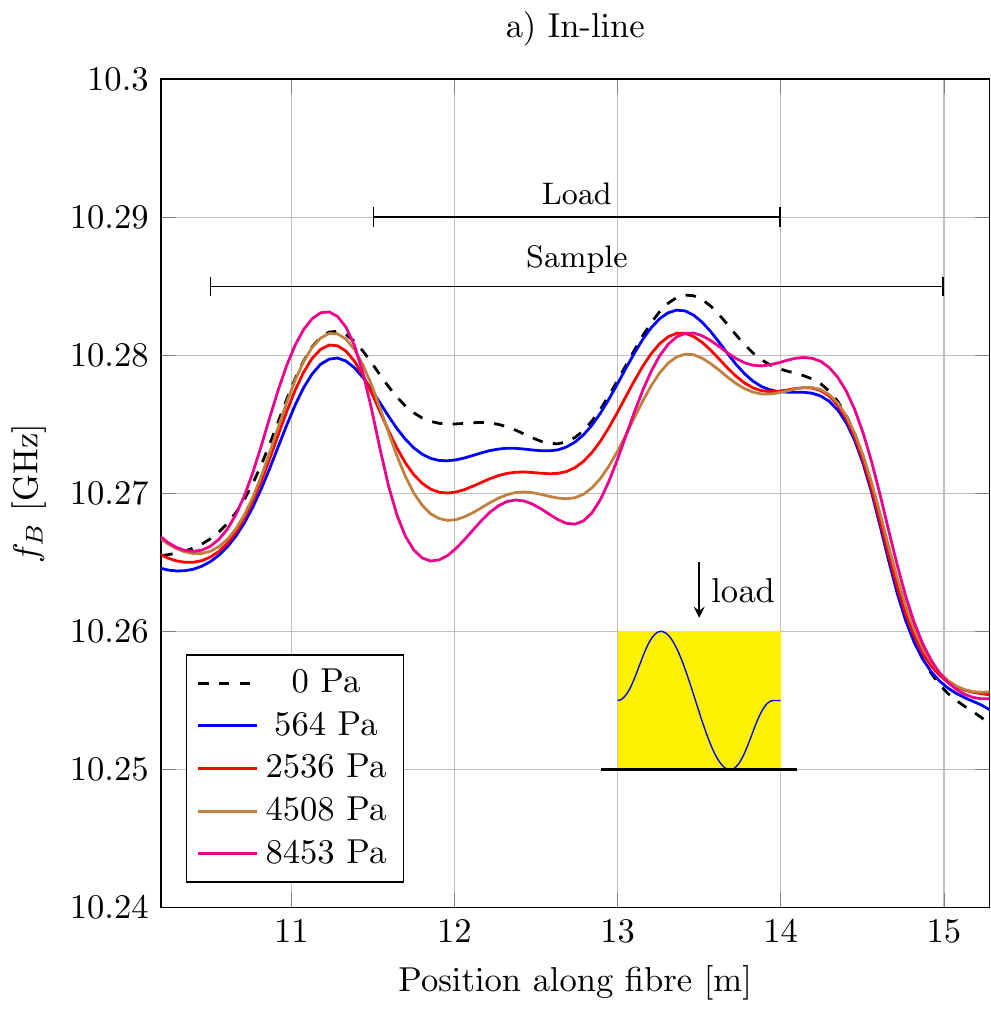}
    \end{subfigure}%
    \begin{subfigure}{.5\textwidth}
        \centering
        \includegraphics[width=0.9\textwidth]{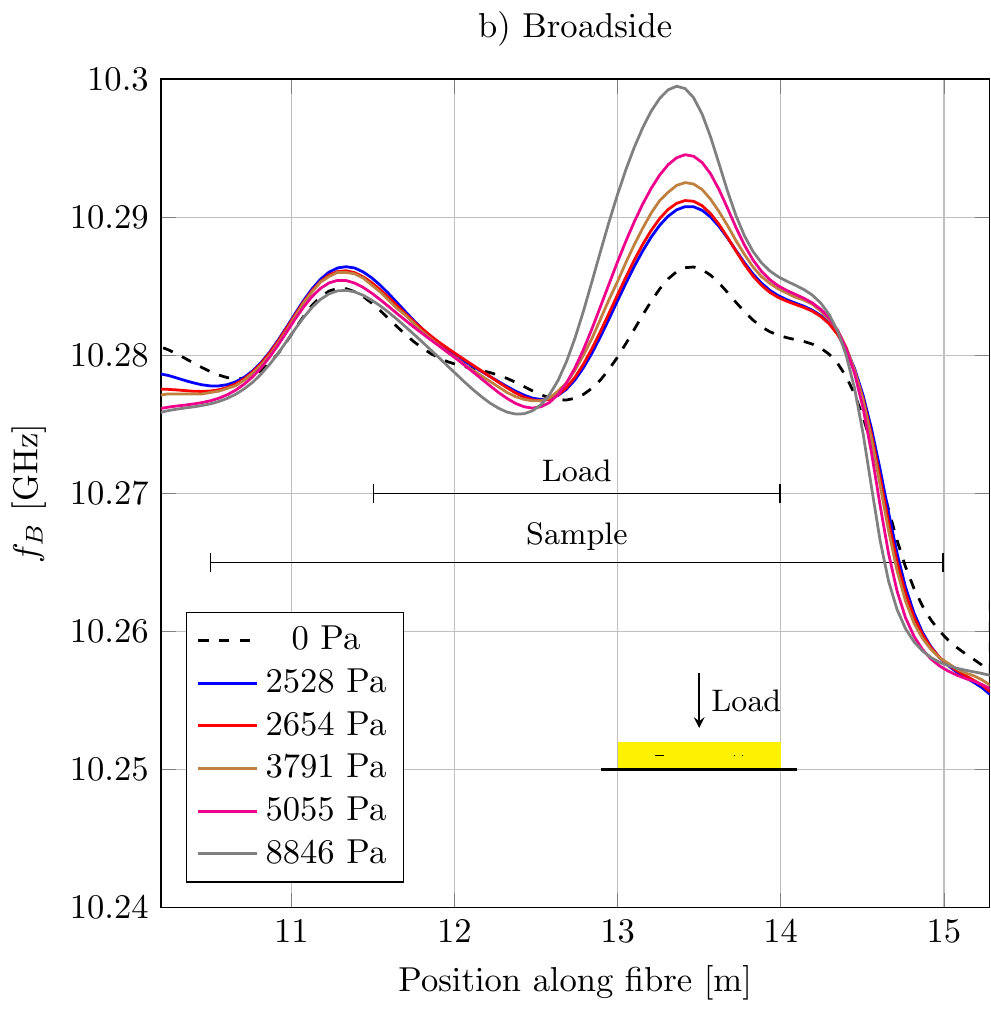}
    \end{subfigure}%
\caption{\textit{Profiles of Brillouin Frequency Shift for in-line (a) and broadside (b) loads. }}
\label{fig:bsf_profiles}
\end{figure}

In figure \ref{fig:bsf_profiles} the results of these measurements are shown.
The profiles of the measured Brillouin frequency shift ($f_B$) for both orientations show opposite behaviour; that is, increasing stress in the in-line orientation caused a decrease in frequency shift as shown in figure \ref{fig:bsf_profiles}a, whereas increasing stresses in the broadside orientation causes an increase in frequency shift as shown in figure \ref{fig:bsf_profiles}b. This is consistent with what was expected.

\begin{figure}[H]

    \includegraphics[width=\textwidth]{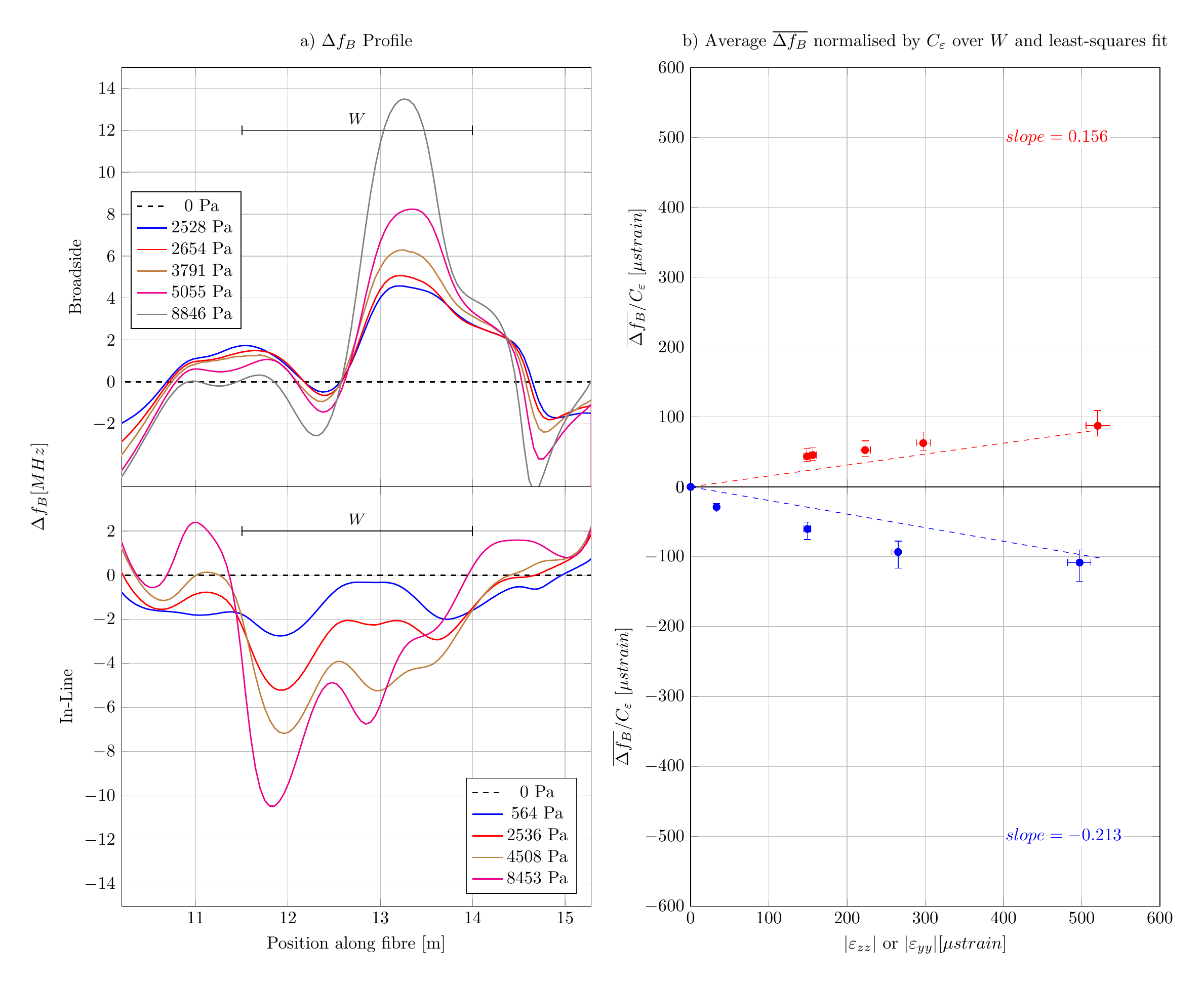}
     \caption{\textit{  Analysis of differences in Brillouin Frequency Shift. a) Differences in profiles compared to reference (zero-load) case, with analysis window $W$ indicated.  Note that due to the limited spatial resolution, the window is slightly extended by $25 cm$ at both sides to average out the effects associate with this limitation. b) $\overline{\Delta f_B}/C_\varepsilon$ in $\mu \varepsilon$ (dots) as a function of applied strain, where the bar above $\Delta f_B$ means the average over window $W$. A least-squares fit for a dashed line that goes through the origin is shown, with the slopes, i.e., sensitivities, given. Note that an offset in the intercept was applied in the least-squares fit to start from the origin.  
     }
     \label{fig:BFS_Analysis}}
     
\end{figure}

To quantify the sensitivity to strain, the difference in frequency shift $f_B$ for each weight relative to a reference measurement (i.e. when no weight is applied) is shown in figure \ref{fig:BFS_Analysis}a.  The average value for each change in frequency shift ($\Delta f_B$) is determined and plotted against the absolute value of the applied strain, $\varepsilon_{ii}$. The applied strains are derived using $\varepsilon_{ii} = \sigma_{ii}/E_c$ for the average value of Young's modulus ($E_c = 17$ MPa). Note that the variations in Young's modulus (i.e. table \ref{tab:matProp}) are incorporated in the error bar for the horizontal axis by calculating the lower and upper bounds of the applied strains  (fig. \ref{fig:BFS_Analysis}b).

As can already be seen in figure \ref{fig:bsf_profiles}, the change in Brillouin frequency shift is not constant along the profile, while the load was applied uniformly. To accommodate these variations and the limited spatial resolution, an average $\overline{\Delta f_B}$ of the differences in the frequency-shift normalised by the strain coefficient $C_\varepsilon$ is calculated over a window $W$ corresponding to the load position (fig. \ref{fig:BFS_Analysis}). This should be equivalent to the fibre strain $\varepsilon_f$ as described by equation \ref{eq:strain}. The strain coefficient $C_\varepsilon$ used is $50$ MHz/1000 $\mu \varepsilon $. Note that $C_\varepsilon$ was not measured for the embedded fibre, and typical corner values of $40$ MHz/1000 $\mu \varepsilon$ and $60$ MHz/1000 $\mu \varepsilon$ are used to evaluate the vertical error bars. 

When looking at the points in figure \ref{fig:BFS_Analysis}b, a main linear trend can be observed. We fitted a line to go through these points while shifting the intercept go through the origin. The slopes of these lines, i.e., the sensitivities, are determined and given in figure \ref{fig:BFS_Analysis}b.
On top of the linear trend, we can see some non-linear behaviour of $\overline{\Delta f_B}/C_{\varepsilon}$ as a function of applied strain. These variations can already be observed in the plots of the $\Delta f_B$ profiles for different loads in figure \ref{fig:BFS_Analysis}a.
These variations are suspected to be mainly caused by two sources, the first one being related to the shape of the sinusoidal fibre and the spatial resolution, where for every interval, the sinusoidal shape is not the same. The other source could be the difference in thickness along the strip or the non-uniform pre-straining of the fibre during the manufacturing process resulting in imperfectly distributed deformation. 

\section{Modelling versus Experiments}
\begin{figure}[H]
    \centering
    \includegraphics[width=0.7\textwidth]{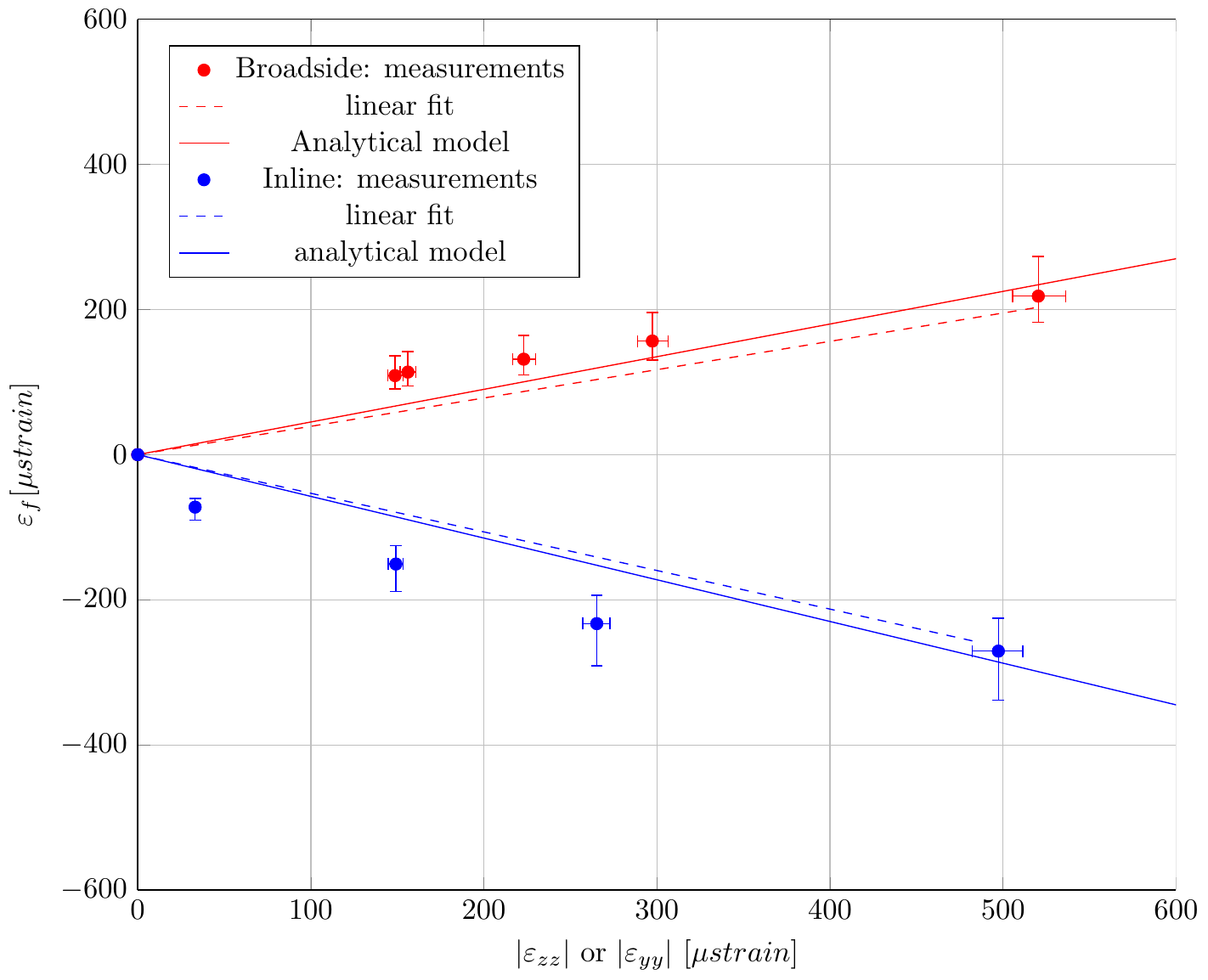}
     \caption{\textit{Comparison between measured ($=\overline{\Delta f_B}/(C_{\varepsilon}\xi)$) and modelled $\varepsilon_f$ for in-line and broadside orientations.}
     \label{fig:df_vs_eps}}
     
\end{figure}

The theoretical fibre strain is calculated using equation \ref{eq:ec}. Ideally, this should give a fit to the data for a fibre made of Conathane\textsuperscript{\textregistered}. However, when the fibre strain was calculated, a scaling factor was needed to have the theoretical fibre strain in the same range as the measured fibre strains.  This scaling factor obtained from the broadside fitting was then also used for scaling the model to fit the measurement of the in-line deformation . So a factor $\xi$ was introduced as:
\begin{equation}\label{eq:eps_f}
    \varepsilon_{f,\mbox{obs}} = \xi \, \varepsilon_{f,\mbox{model}} .
\end{equation}
where $\varepsilon_{f,\mbox{obs}}$ is the observed strain ($\overline{\Delta f_B} /C_{\varepsilon}$), and $\varepsilon_{f,\mbox{model}}$ the modelled strain.
We suspect that this scaling factor is related to a coupling mechanism between the fibre and the embedding cable material due to the significant difference in elastic properties between the two, as shown in table \ref{tab:matProp}.  The scaling factor $\xi$ found to fit the measurements is 0.4. Using such a value, figure \ref{fig:df_vs_eps} shows that a good agreement between the measurements, when divided by $\xi$,  and the model is made. 

Even though the exact value of $C_\varepsilon$ is unknown, its impact on the variation of equivalent strain values is quite limited as illustrated by the small vertical error bars in figure \ref{fig:df_vs_eps}. To compare the sensitivities in the different orientations, the ratio of the different orientations would provide a suitable measure. With such a ratio, the coupling factor $\xi$ as introduced earlier in equation \ref{eq:eps_f} is used to calibrate the the measurements.  We find that the modelled ratio is 1.28 and the ratio based on the least-squares fits to the calibrated data is 1.33. The small variation between the two ratio's could be related to the variations in the $\Delta f_B$ measurements as discussed earlier.

It is safe to say that that the strip is sensitive to both orientations, although with opposite signs. The sensitivity to the broadside load is seen as mainly due to the Poisson effect of the Conathane\textsuperscript{\textregistered}. As depicted in figure \ref{fig:BFS_Analysis}, for the analysis window ($W$), the results show a slight increased sensitivity in the in-line direction compared to broadside direction. 
This is attributed to the high Poisson's ratio of Conathane\textsuperscript{\textregistered} (i.e. $\nu_c \approx 0.45$); that is, when the loads are applied on the broadside or $y$-direction, the fibre is extended in both the $x$- and $z$-direction, as was shown in figure \ref{fig:sinus_BS}. In contrast, when applying the loads on the in-line or $z$-direction,  the fibre will experience shortening in the $z$-direction and extension in the $x$-direction, as was shown in figure \ref{fig:sinus_IL}; the latter is due to Poisson's ratio. Therefore these two effects work in opposite directions to each other. Clearly, the shortening in $z$-direction is dominating, and we see that in the results too.

As can be seen the model pretty well describes the behaviour we observe from the measurements with the Conathane\textsuperscript{\textregistered} sample, apart from the factor $\xi$ that needed to be introduced. Quantitatively, there are some uncertain issues: 
\begin{itemize}
    \item The broadside sensitivity is seen as mainly due to the Poisson effect of the embedding material, i.e., the strip/cable of Conathane\textsuperscript{\textregistered}. What would be ideal in such a situation is that the embedding material would have a Poisson ratio of zero for a broadside load.
    \item In the experiments, a straight fibre in the strip was not included while this would have been insightful on the deformation in the $x$-direction. It also would have been useful to better understand the coefficient $\xi$ that needed to be introduced in the model to fit the data (or vice-versa). 
    \item For the Conathane\textsuperscript{\textregistered}, a separate small experiment took place to determine Young's modulus and Poisson's ratio, but those ones were measured at relatively large strains, in the order of 10 m$\varepsilon$ rather than 0.1 m$\varepsilon$. Determining Young's modulus and Poisson's ratio of Conathane\textsuperscript{\textregistered} at lower strain values would give more realistic values. 
    \item A more reliable estimate of the strain coefficient for the fibre, i.e. $C_{\varepsilon}$ (see eq.\ref{eq:strain}), would make the conversion more reliable too; however, this is not possible with our current set-up. 
\end{itemize}

\section{Conclusion}
The concept of a shaped fibre was adapted in this study to enhance the directional sensitivity. Some relatively simple analytical modelling and Brillouin-scattering-type experiments were carried out on a 2-m polyurethane (Conathane\textsuperscript{\textregistered}) strip embedding the shaped fibre. To examine its sensitivity to in-line and broadside deformations, loads were applied in these orientations, where an opposite behaviour was observed: a negative strain with increasing in-line loads and a positive strain with increasing loads in broadside orientation, with a slight difference in the sensitivity in these two orientations.

Our developed model describes the observations quantitatively, but some uncertainties are limiting a quantitative estimation of the sensitivity. These uncertainties pertain to the strain coefficient $C_{\varepsilon}$ of the fibre as well as low-strain measurements of the elastic properties of the embedding material $E_c$ and $\nu_c$. On the positive side, the model highlights the importance of the Poisson's ratio in the broadside orientation and can be also used to optimise the sensitivity in the in-line orientation based on the geometrical parameters $A$ and $\Lambda$. 
Although polyurethane material (in our case  Conathane\textsuperscript{\textregistered}) is not directly suitable for unidirectional strain measurements using a sinusoidal fibre, this study shows that the embedding material is dominating the behaviour of the total material, and it is therefore crucial to take this into account when fibres are embedded in cables as a strain-sensing device.

\section*{Acknowledgement}
This research has received funding from the European Research Council (ERC) under the European Union’s Horizon 2020 research and innovation program (grant no. 742703). We would like to thank Karel Heller and Jens van Berg for the assistance with constructing the laboratory set-up, Xuehui Zhang, for the help with the sensing instrument and Richard Bakker for his support in estimating the elastic properties of the material.

\bibliography{references}
\bibliographystyle{apalike}

\end{document}